\documentclass[12pt]{iopart}

\usepackage{iopams}
\usepackage{graphicx}  %este lo agregu\'e pero est\'a sugerido en guidelines

\begin{document}

\title[The effects of uneven load sharing]{Tug-of-war of molecular motors: the effects of uneven load sharing}.
%\title[load sharing in models for tug of war]{Equal vs. unequal sharing load sharing in models for tug of war}
%\title[load sharing in models for tug of war]{Equal vs. unequal load sharing in models for bidirectional transport in microtubules}
\author{Sebasti\'an Bouzat$^{1}$ and Fernando Falo$^{2}$}

\address{1 Consejo Nacional de Investigaciones Cient\'{\i}ficas y T\'ecnicas, Centro At\'omico Bariloche (CNEA), (8400) Bariloche, Argentina.\\
2  Dpto de F\'{\i}sica de la Materia Condensada and BIFI, Universidad de Zaragoza,
50009 Zaragoza, Spain.}

\ead{bouzat@cab.cnea.gov.ar}

\begin{abstract}
We analyze theoretically the problem of cargo transport along microtubules by motors of two species
with opposite polarities. We consider two different one-dimensional models previously developed in
the literature. On the one hand, a quite widespread model which assumes equal force sharing, here
referred to as mean field model (MFM). On the other hand, a stochastic model (SM) which considers
individual motor-cargo links. We find that in generic situations the MFM predicts larger cargo mean
velocity, smaller mean run time and less frequent reversions than the SM. These phenomena are found
to be consequences of the load sharing assumptions and can be interpreted in terms the probabilities
of the different motility states. We also explore the influence of the viscosity in both models and
the role of the stiffness of the motor-cargo links within the SM. Our results show that the mean
cargo velocity is independent of the stiffness while the mean run time decreases with such a
parameter. We explore the case of symmetric forward and backward motors considering kinesin-1
parameters, and the problem of transport by kinesin-1 and cytoplasmic dyneins considering two
different sets of parameters previously proposed for dyneins.
\end{abstract}

\vspace{2pc}

\submitto{\PB}
\maketitle

%\twocolumn %Esto lo agrego para probar el formato, pero no s\'e si es el de IOP

\section{Introduction}

Transport of cargo driven by multiple molecular motors along microtubules has become a very active
subject of research because of its relevance for many cellular functions
\cite{Howard,Schliwa03,Goldstein,gross2004}. In recent years, a myriad of experiments and models have
attempted to understand the way in which motors work together
\cite{gross2004,welte04,fisher07,lipowskyPhysE,grossCB2008,lipowsky09}, and, still, there are many
fundamental details which remain unclear and deserve further research, most particularly for the case
of bidirectional transport by two motor species.

The complexity of the multiple motor systems and the difficulties for controlling the experiments are
often quite important so that performing the connection between models and experiments must be done
carefully. Models involve always many parameters, including for instance detachment and attachment
rates, stall forces, motor stiffness and viscosity of the media. Usually, many of these parameters
are a priori not well known in the experiments, and even more fundamental features such as the number
of motors, or whether more than a single species is participating on the transport, remain unclear.
Thus, distinct models may provide different fitting of the experimental data and, consequently,
different interpretations. Moreover, recent in vivo experiments \cite{cellgross2008} have revealed
important differences with in vitro systems. In this context, a detailed knowledge of the
consequences of specific modeling assumptions as well as the comparison of different kinds of models
becomes quite relevant. The aim of this paper is to contribute in these two important aspects.

References \cite{klumpp2005} and \cite{tugofwar} have originated a modeling framework that has
largely contributed to the understanding of transport by several motors. The model introduced in
\cite{klumpp2005} deals with cargo transport by a single class of motors, while in \cite{tugofwar}
the formalism is extended to account for bidirectional transport associated to tug of war between two
motor types with opposite polarities. Assuming certain force--velocity relations, and specific
attachment and detachment probabilities for individual motors, the model enables the calculation of
the probabilities of different motility states characterized by different number of motors, and the
reproduction of trajectories and velocity distributions as well. In a series of papers
\cite{lipowskyPhysE, Muller, ZhangMF, lipowskiBioPJ} the model was further developed and several
effects and transport conditions have been analyzed, providing a deep physical insight on the
problem. An important assumption of the model is that all the motors of the same polarity
simultaneously engaged to the microtubule share equally the load. In real systems, however,
fluctuations of the distances between motor-microtuble binding position and motor-cargo binding
position may lead to non-negligible differences between the forces supported by the different motors
\cite{grossCB2008,Erickson,mogilner,Jamison}. Consequently, the model would eventually fail to
predict exact quantitative results. In reference \cite{mogilner}, the model was referred to as {\em
mean field} due to the equal sharing of load approximation. We will keep such a name throughout this
work.

Several models have gone beyond the mean field approach by considering independent motor-cargo links
for each motor, and incorporating different degree of detail in their description of individual motor
properties \cite{grossCB2008,lipowsky09,mogilner,zgz,ZhangSM}. Although such models generally provide
less instrumental (and less elegant) formulations than the mean field model, and they mostly lack
analytical results, they may be more successful in predicting numerical results for multiple motors
through simulations based on individual motor parameters. In a different but related context, models
in references \cite{prlcampas} and \cite{brugues2009} consider the load applied only to the leading
motor and constitute thus interesting extreme examples of models beyond mean field. Although not
directly connected to our approach for processive motors on microtubules, studies on non-processive
motors \cite{Hexner, Gur} and general ratchet models \cite{Gillo} provide also relevant analysis of
bidirectional motion in many motor systems.

In this paper we investigate bidirectional cargo transport by two opposing teams of processive motors
within two different models. On the one hand, the mean field model. On the other hand, a recently
introduced \cite{zgz} stochastic model which considers independent cargo-motor links for individual
motors, allowing for uneven load sharing. In this way, at the same time that we investigate how cargo
transport depends on the system parameters, we are able to clearly identify the consequences of the
assumption of equal load sharing. Our work follows the spirit of the paper by Kunwar and Mogilner
\cite{mogilner}. There, the authors compared results from both kind of models focussing on the case
of cargo transport by a single team of motors, and provided also an analysis of the velocity
distributions for bidirectional transport. Moreover, they studied the influence of the non
linearities of the force-velocity relations of individual motors.

Our studies focus on analyzing the dependence of cargo transport on the number of motors of each
polarity, the viscous drag, and the stiffness of the motor cargo link, while we do not consider the
influence of additional load forces acting on cargo. In Section 2 we present the models. Section 3
studies the case of equal forward and backward motors. The effects of varying the number of motors to
each side and the influence of viscous drag are analyzed within both models. In section 4 we present
results for bidirectional transport by asymmetric motors considering system parameters compatible
with kinesin-1 and cytoplasmic dynein. Section 5 is devoted to the conclusions.

\section{Models and methods}

As indicated in the introduction, we will consider two different
models for the analysis of cargo transport by multiple motors. The
mean field model (MFM), and our recently proposed stochastic model
(SM). We first introduce the characteristics that are common to
both.

The two models consider the cargo as a point particle which performs
a continuous trajectory $x(t)$ in one dimension. The cargo is linked
to $N_f$ forward motors and $N_b$ backward motors. The first of them
can pull the cargo in the positive direction while the second can
pull it in the negative one. At a given time, the number of forward and
backward motors engaged to the microtubule are respectively
$n_f(t)\leq N_f$ and $n_b(t)\leq N_b$. Each engaged motor $i$,
detaches from the microtubule with a probability per time unit given
by $\epsilon \exp(|f_i|/F_d)$. Here $f_i$ is the instantaneous force
exerted by motor $i$ on the cargo, $\epsilon$ is the reference
zero-load detachment rate and $F_d>0$ is the detachment force. We
will call $\epsilon_f, \epsilon_b, F_{df}$ and $F_{db}$ the
corresponding parameters for forward and backward motors.
Conversely, a detached motor engages to the microtubule with rate
$\Pi_f$ or $\Pi_b$, according to its type.

When loaded with a force $f_i>0$ (considered positive if exerted
against the polarity of the motor), the motor $i$ advances with
velocity
\begin{equation}
\label{vi} v_i=\left\{
\begin{array}{lc}
v_0(1-f_i/F_s) &{\rm for} f_i\leq F_s \\
v_1(1-f_i/F_s) &{\rm for} f_i > F_s.
\end{array}\right.
\end{equation}
Here $F_s>0$ is the stall force, $v_0$ the zero-load velocity and $v_1$ a reference backward
velocity. Considering both motor species we have the system parameters $F_{sf}, F_{sb}, v_{0f},
v_{0b}, v_{1f}$ and $v_{1b}$. The linear force-velocity relation for single motors of Eq. (\ref{vi})
is a natural choice for comparing the SM and MFM since it is used in most works on the MFM
\cite{lipowskyPhysE,tugofwar,Muller} and is also a common assumption in other theoretical models
\cite{mogilner,nedelec,wollman}. Studies in \cite{mogilner} suggest that the consideration of a
general non-linear relations would lead to no relevant qualitative changes in the results. It is
important to mention that, while Eq.(\ref{vi}) is taken as instantaneously exact in the MFM, within
the SM it is only valid in terms of time averages, i.e. $v_i$ is the mean velocity of a motor subject
to a constant force $f_i$.

The way to compute the forces $f_i$ and the cargo motion depends on
the model as we explain in the following subsections.

\subsection{Cargo dynamics in the mean field model}

The MFM \cite{tugofwar} assumes that all the motors (backward and
forward) move with the same velocity than cargo at any time, and
that motors of the same polarity share the force equally. It also
assumes the total force acting on the cargo vanishes at almost any
time (it has discontinuities at the times at which the number of
engaged motors changes). With such hypothesis, by performing a force
balance and using the force-velocity relation for single motors of
Eq.(\ref{vi}), it is possible to obtain the cargo velocity as a
function of the numbers of engaged motors $n_f$ and $n_b$
\cite{tugofwar}. We thus have a discrete set of allowed cargo
velocities $v(n_f,n_b)$, corresponding to the different motility
states $(n_f,n_b)$ considering $n_f=0,1,...,N_f$, $n_b=0,1,...,N_b$.
The model can be implemented through two main different methods.
First, by means of a master equation which allows to compute
stationary probabilities $P(n_f,n_b)$ and, thus, velocity
distributions $P(v)$. And, second, by means of a Gillespie algorithm
\cite{tugofwar,gillespie} which allows to compute cargo
trajectories. This latter numerical scheme determines the temporal
evolution of the system by ruling the transitions between different
motility states taking into account the attachment and detachment
probabilities. During each time interval between two transitions,
the cargo velocity is assumed to be constant and equal to the
corresponding value $v(n_f,n_b)$.

For two motors species, the model was first introduced in
\cite{tugofwar} without considering any external force. Then, in
\cite{lipowskiBioPJ} it was generalized to include the cases of
viscous environments and non vanishing load forces. This is done by
modifying appropriately the force balance, but without changing any
of the model hypothesis mentioned before.

As part of our studies we have implemented both the master equation
and Gillespie formulations of the model without external forces
(although we will only show results from the latter), and also the
Gillespie method considering non negligible viscous drag. We have
checked that our results correctly reproduce some selected ones from
references \cite{tugofwar} and \cite{lipowskiBioPJ}. In all cases we
have assumed the linear force-velocity profile of Eq.(\ref{vi}).

\subsection{Stochastic model}

As other models in the literature \cite{grossCB2008,mogilner}, the stochastic model introduced in
\cite{zgz} considers a Langevin dynamics for cargo motion, a discrete-steps stochastic dynamics for
individual motors and cargo-motor links described by non-linear springs.  The Langevin equation for
cargo is
\begin{equation}
\gamma \dot x_c=\sum_i f_i +\xi(t).
\end{equation}
Here $\gamma$ is the viscous drag, $f_i$ ($i=1,...,N=N_f+N_b$) the force exerted by the $i$-th motor
and $\xi(t)$ the white thermal noise. The viscous drag is defined through the Stokes relation
$\gamma=6 \pi \eta r$ \cite{grossCB2008,mogilner}, where $\eta$ is the viscosity of the medium and
$r$ the radius of the cargo for which we consider $r=500 nm$ throughout the paper. The thermal noise
satisfies  $\langle \xi(t)  \rangle= 0$ and the correlation formula $\langle \xi(t_1) \xi(t_2) \rangle=2 D \delta(t_1-t_2)$
\cite{vankampen}. Here $\langle \, \rangle$ represents ensemble average, $\delta(t)$ is the Dirac
Delta and $D$ the diffusion coefficient satisfying the fluctuation-dissipation relation $D=k_B T /
\gamma$ \cite{vankampen}, with $k_B$ the Boltzmann constant and $T$ the temperature. In all our
calculations we consider $T=300 K$.

Each motor is modeled as a particle that can occupy discrete positions separated by $\Delta x=8nm$
along the same spatial coordinate used for the cargo. Its dynamics is governed by a Monte Carlo
algorithm \cite{zgz} that rules the elementary processes of step forward, step back, detachment, and
attachment. At each time step of duration $dt$, an engaged forward motor has a probability
$p_{jump}=dt/\tau_D(F)$ of performing an $8nm$ step, which may be forward (right) with probability
$P_r(F)=\left[R(F)/(1+R(F))\right]$ or backward (left) with probability
$P_l(F)=\left[1/(1+R(F))\right]$. Here $\tau_D(F)$ is the dwell time \cite{cross2005,block2000} and
$R(F)$ is the forward-backward ratio of jumps \cite{cross2005,block2000,hyeon09}. The resulting mean
velocity for a single forward motor with constant load $F$ is $v(F)=\Delta x
(P_r(F)-P_l(F))/\tau_D(F)$. In \cite{zgz} the model was developed assuming certain specific formulas
for $R(F)$ and $\tau_D(F)$ based on experimental data for kinesin-1, while $v(F)$ was left as free.
Here, in order to compare results with the MFM, we consider $v(F)$ as a known relation instead of
$\tau_D(F)$. Then, the value of $\tau_D(F)$ entering in the algorithm is determined by inverting the
corresponding formulas. For the $R(F)$ we consider the experimentally based
\cite{cross2005,block2000,hyeon09} formula $R(F)=A \exp(-\log(A) |F|/F_s)$, with $A=1000$ and $F_s$
the before mentioned single motor stall force that leads to $P_r=P_l$. For the backward motors we
consider the same single motor model but interchanging right and left. The forces $f_i$ are computed
assuming the cargo is linked to each motor by a non linear spring \cite{grossCB2008,mogilner} which
produces only attractive interactions, and only for distances larger than a critical one. Let us call
$x_i$ the position of motor $i$ and $\Delta_i=x_i-x_c$. We define $f_i=k(\Delta_i-x_0)$ for
$\Delta_i\ge x_0$, $f_i=0$ for $-x_0<\Delta_i< x_0$ and $f_i=k(\Delta_i+x_0)$ for $\Delta_i\le -x_0$,
with $x_0=110\, nm$ \cite{grossCB2008,mogilner}. Here, $k$ is the stiffness of the motor for which we
consider values $k_f$ and $k_b$ for forward and backward motors respectively. Note that while in
\cite{zgz} we have included volume excluded interaction between motors, here we consider only
interactions mediated through cargo.

The detachment and attachment processes occur according to the
probabilities per time unit indicated at the beginning of this
section. The attachment of detached motors occurs with equal
probability in any of the discrete sites $x_j$ satisfying
$|x_j-x_c|<x_0$.

\subsection{Relevant quantities and numerical simulations}

We study the cargo dynamics within MFM and SM by performing
numerical simulations of the evolution of the system for different
values of the parameters. As initial condition (at time
$t_{ini}\equiv 0$) we consider a random number of motors of each
species engaged on the microtubule. Each realization finishes when all
the motors are detached (at time referred to as $t_{end}$). The
numerical simulations of the SM are performed as explained in
\cite{zgz} using time steps between $dt=2 \times10^{-5}s$ and $dt=3
\times10^{-7}s$ depending on the value of $\gamma$. For the MFM we
use our implementation of the Gillespie algorithm explained in
\cite{tugofwar,gillespie}.

In order to characterize the long-time properties of cargo dynamics
we compute the following quantities.
\begin{itemize}
\item {\em Cargo mean velocity}. Defined as the average over realizations
of the ratio $(x_{end}-x_{ini})/(t_{end}-t_{ini})$.
\item {\em Run length.} Defined as the average over realizations of $(x_{end}-x_{ini}).$
\item {\em Run time.} Denoted as $\tau_r$, equal to the average of $(t_{end}-t_{ini}).$
\end{itemize}

Concerning the analysis of the dynamical properties during forward and backward stages of the motion,
we compute the {\em mean forward run length} ($r_f$) and the {\em mean backward run length} ($r_b$).
We define them as the average distance traveled by the cargo during the time intervals at which
$n_f>n_b$ and $n_f<n_b$ respectively. Note that the association of forward (backward) motion of the
cargo with $n_f>n_b$ ($n_f<n_b$) makes sense only for symmetric motors (i.e. equal parameters for
motors of both polarities). For asymmetric motors, the characterization of forward and backward
stages of motion demands a signal analysis of the trajectories including filtering and appropriate
definitions of switching points and forward and backward runs, as it is usually done in experimental
works \cite{filtros,valeria}. Such kind of studies is out of the scope of the present work.

Note that, for the MFM with symmetric motors, any definition of forward and backward run lengths
based on signal analysis of trajectories would lead exactly to the same results as our definitions
based on $n_f$ and $n_b$. This is because, in such a case, the condition $n_f>n_b$ or $n_f<n_b$
determine the direction of motion. In contrast, within the SM, the exact coincidence between both
kind of definitions cannot be ensured, since for very short times we could have forward (backward)
motion of cargo with $n_f<n_b$ ($n_b<n_f$). Nevertheless, the differences are expected to be small.
In any case, the definitions in terms of $n_f$ and $n_b$ are relevant by themselves for our
theoretical analysis.

\section{Results for symmetric motors}

First we analyze the results for both models considering equal parameters for forward and backward
motors. For shortness, we speak of equal or symmetric motors. Except when specially stated, we
consider single motor parameters compatible with kinesin-1 \cite{tugofwar} for both models:
$F_{s}=6pN, v_{0}=1000 nm/s, v_{1}=6nm/s, \epsilon=1/s, F_{d}=3.18 pN$ and $\Pi_f=5/s$. We left the
numbers of motors $N_f$ and $N_b$, and the viscosity as free parameters. For the SM, except when
indicated, we consider the parameter $k=0.32 pN/nm$ usually taken as reference value for kinesin-1
\cite{mogilner}.

\subsection{Trajectories}

\begin{figure}[h!]
\centering
\resizebox{\columnwidth}{!}{\includegraphics{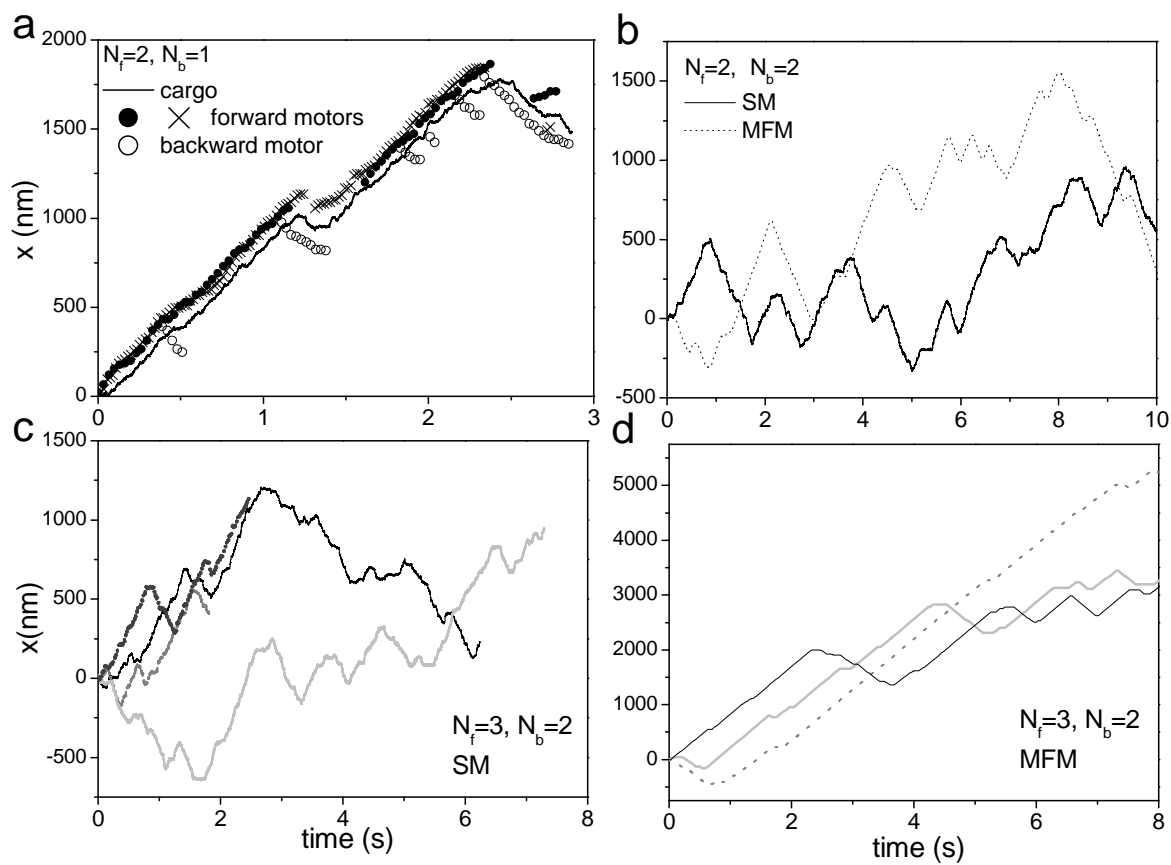}}
\caption{\label{figtray} Trajectories. a) Typical cargo and motors
trajectories for the SM considering $N_f=2, N_b=1$. b) Cargo
trajectories for $N_f=N_b=2$ for SM and MFM. c) Several different
cargo trajectories for $N_f=3$ and $N_b=2$ for SM. d) Ibid c for
MFM. In all the cases the viscosity considered is 100 times the
water viscosity.}
\end{figure}

As a first step in our study, we glance at the trajectories within
both models. Figure \ref{figtray}.a shows cargo and motors
trajectories for a system with $N_f=2$ and $N_b=1$ computed using
the SM. Regions of tug of war leading to pauses and reversions of
the cargo motion can be appreciated. In figure \ref{figtray}.b we
show MFM and SM cargo trajectories for $N_f=N_b=2$. At first glance
we see that both models produce similar trajectories for such
parameters. Thus, we can expect that this may lead to compatible
results for ensemble averaged quantities. In contrast, results in
figures \ref{figtray}.c and \ref{figtray}.d indicate us that, in the
case $N_f=3, N_b=2$ both models predict very different results even
at the level of single trajectories. Thus, depending on the
parameters we may expect that the two models give results which may
be statistically equivalent or not.

\subsection{Results for negligible viscous drag}

Now we begin our systematic analysis of both models focussing on the
behavior of the cargo mean velocity, run length and run time. We
analyze first the dynamics for negligible viscous drag. To do so, we
consider the MFM without viscous drag \cite{tugofwar}, and the SM
with a very small value of $\gamma$, so that the system is
essentially at the zero viscosity limit. Actually, we use
$\gamma=9.42\,10^{-6} pN s/nm$, calculated using the Stokes formula
\cite{grossCB2008,mogilner} with water viscosity and a radius of the
cargo equal to $0.5 \mu m$. Note that for such a value of $\gamma$,
even if we consider a fast cargo velocity of $10^3 nm/s$ we get a
viscous drag of order $10^{-2} pN$ which is quite small compared to
the typical forces on the scale of $1 pN$ involved in motor
dynamics.

\begin{figure}[h!]
\centering
\resizebox{\columnwidth}{!}{\includegraphics{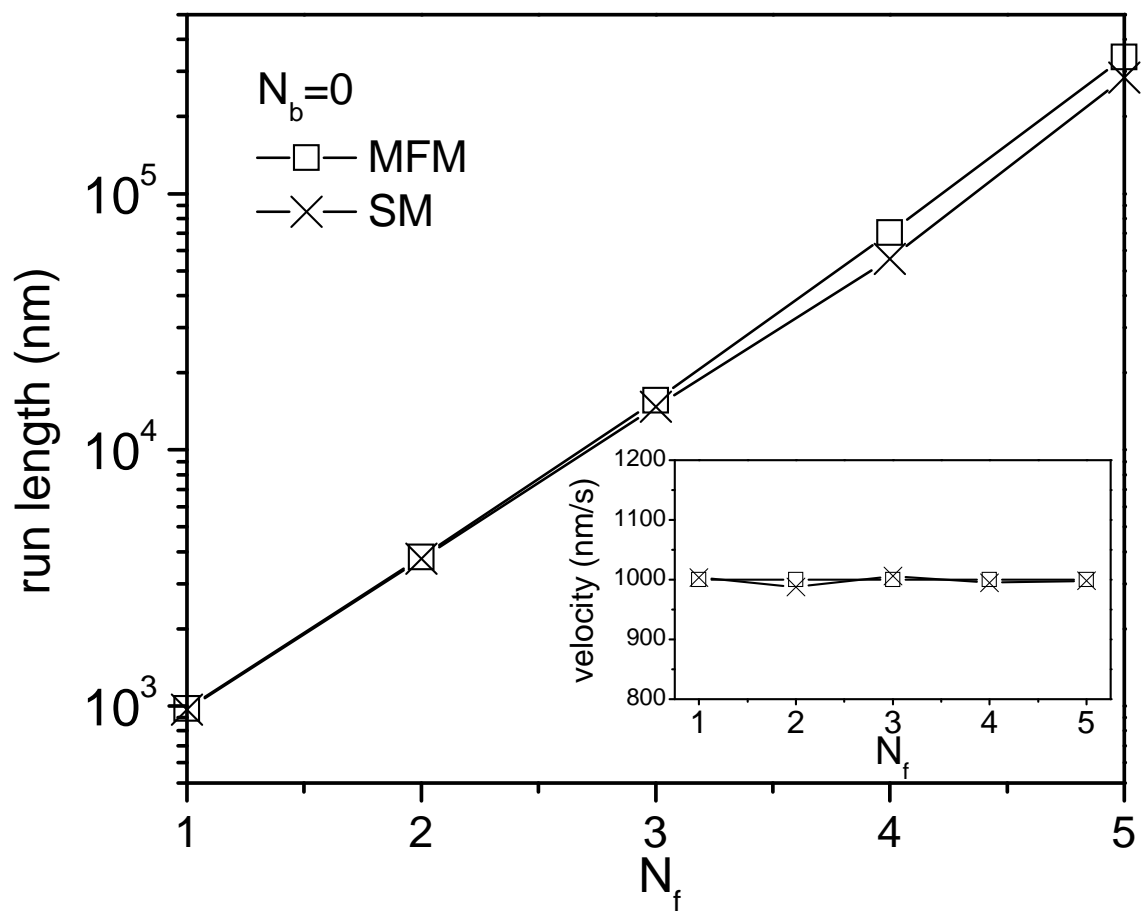}}
\caption{\label{figsolofow} Results for a single team of motors
($N_b=0$). Run length and cargo mean velocity (inset) as functions
of $N_f$ for SM and MFM considering water viscosity.}
\end{figure}

In figure \ref{figsolofow} we study the case of a single species of
motors considered with forward polarity. We plot the run length and
the velocity as functions of the number of motors. The results are
already well known from a number of previous works: the velocity is
independent of the number of motors (for negligible viscosity),
while the run length grows exponentially. Our contribution here is
to show that both models agree in their numerical results. As the
analysis of single trajectories suggest and we will shortly confirm,
this is not always the case when we consider two species of motors.

\begin{figure}[h!]
\centering
\resizebox{\columnwidth}{!}{\includegraphics{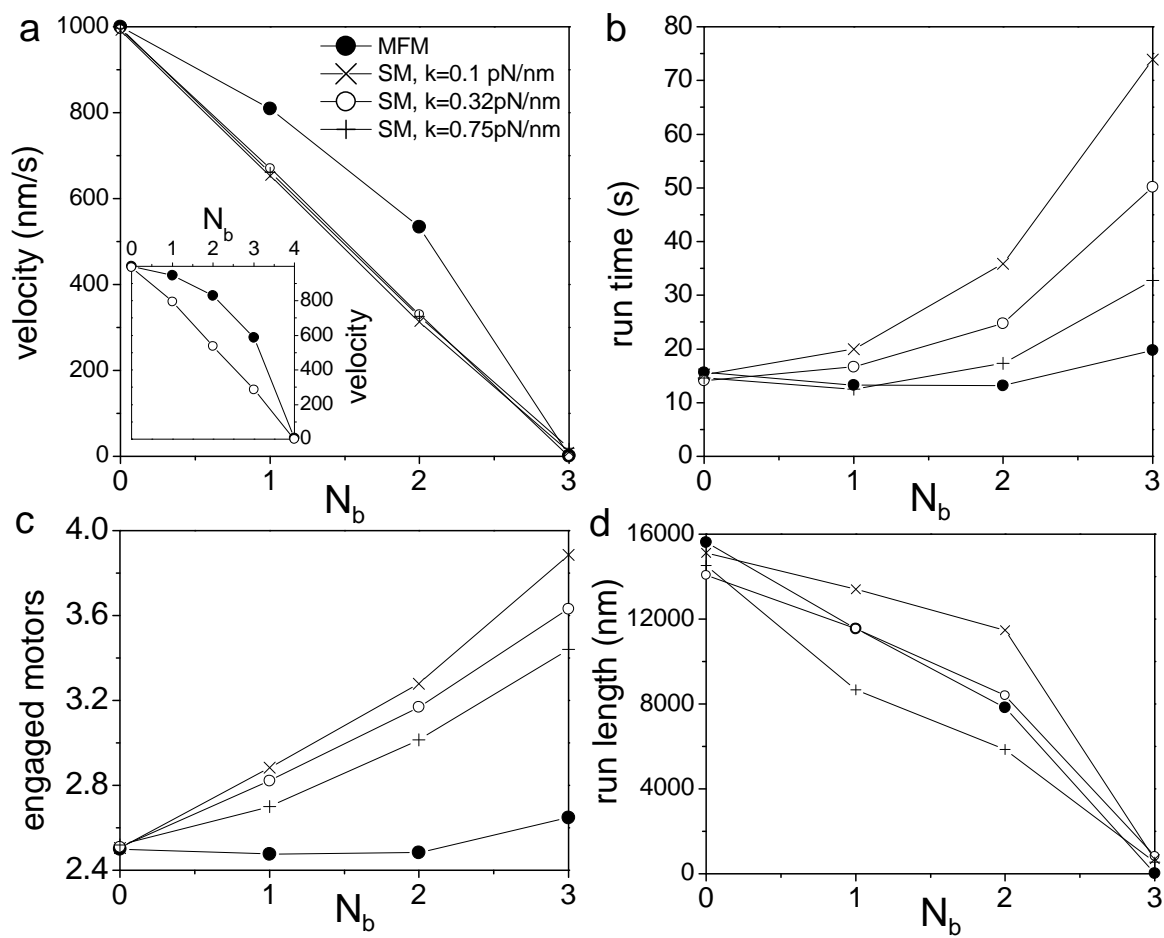}}
\caption{\label{vyrN3} Cargo mean velocity (a), mean run time (b),
mean number of engaged motors (c) and run length (d) as functions of
$N_b$ for systems with $N_f=3$. Results for mean field and
stochastic models under conditions of negligible viscous drag. In
all the cases we consider kinesin-1 parameters both for forward and
backward motors. The inset in panel (a) shows velocity results for
$N_f=4$ whose analogy with those for $N_f=3$ indicates that the
latter case (studied throughout the paper) is typical.}
\end{figure}

In figure \ref{vyrN3}.a we show results for the cargo velocity as a function of the number of
backward motors $N_b$ for fixed $N_f=3$ in the symmetrical case. It can be seen that both models
coincide only for $N_b=0$ and $N_b=N_f$, while for intermediate values of $N_b$ the MFM predicts
considerably larger velocities. These differences are a consequence of the load sharing hypothesis.
Note that, while in the MFM all engaged forward motors contributes equally to pulling the cargo, in
the SM only those motors which are instantaneously beyond the limit distance $x_0$ from cargo exert
non vanishing forces. Hence, each of such {\em pulling motors} are more loaded than motors in the MFM
and, thus, their velocity is smaller. Clearly, this causes a smaller cargo velocity, since cargo
velocity is essentially controlled by such leading motors. It is interesting to realize that for
the SM we obtain the simple linear behavior $v=v_{0f} (N_f-N_b)/N_f$, regardless the value of the
motor stiffness $k$. This demonstrates a certain degree of robustness of the motor team performance
independently of the stalk stiffness. However, as we will see, other relevant quantities do depend on
$k$. In figure \ref{vyrN3}.b we show the run time $\tau_r$ as a function of $N_b$ for the same system
as in figure \ref{vyrN3}.a. We see that, within the SM, $\tau_r$ increases with $N_b$ and decreases
with $k$. Except for vanishing $N_b$, the SM predicts sensibly larger values of $\tau_r$ than the
MFM. Note that, while the cargo velocity is controlled by the pulling motors, the run time is
expected to be essentially determined by the total number of engaged motors, regardless its polarity.
This is because forward and backward motors contribute equally to linking the cargo to the
microtubule (at least for symmetric motors). The relevance of differentiating between engaged and
pulling motors was discussed in \cite{zgz} when analyzing transport by a single species against an
external load. The relation between run time and total number of engaged motors becomes evident with
the results in figure \ref{vyrN3}.c, where we show the mean number of engaged motors as function of
$N_b$ for the same systems in \ref{vyrN3}.a and \ref{vyrN3}.b. The parallelism between curves in
\ref{vyrN3}.b and \ref{vyrN3}.c is apparent. The total number of engaged motors increases with $N_b$,
decreases when passing from MFM to SM and decreases with $k$ within SM. The causes of these
behaviours will be explained later when studying the probabilities of the different motility states.
In figure \ref{vyrN3}.d we show the run length as a function of $N_b$ for the same parameters as
those in figure \ref{vyrN3}.a. and \ref{vyrN3}.b. As expected, the run length decreases with $N_b$ in
both models following the decrease of the mean velocity. Within the SM, the decrease of the run
length with $k$ can be associated to that of the run time and to the invariance of the mean velocity.
This seems compatible with a factorization of the mean values. Interestingly, the results for the MFM
are similar to those for the SM with $k=0.32 pN/nm$. However, this seems to be due to a compensation
between the decrease of the velocity and the increase of $\tau_r$ when passing from the MFM to the
SM.

\begin{figure}[h!]
\centering
\resizebox{\columnwidth}{!}{\includegraphics{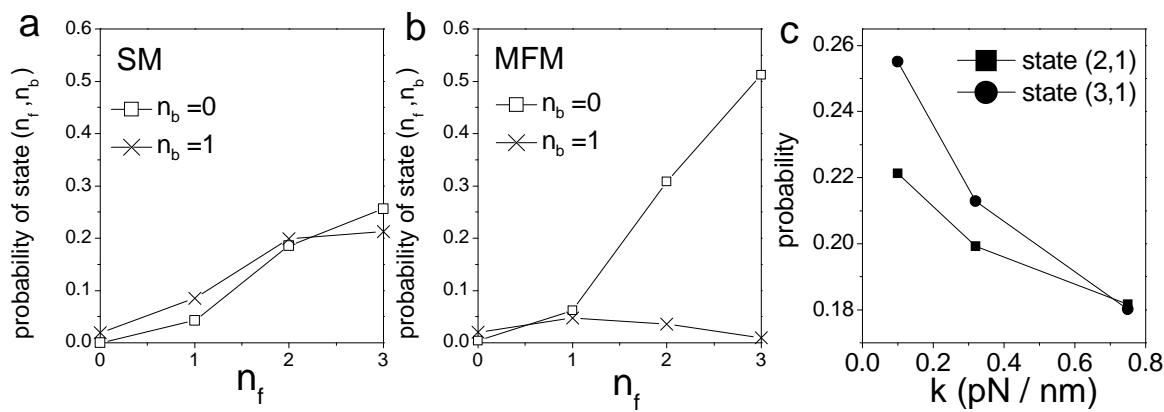}}
\caption{\label{pnfnb} Probability of the states $(n_f,n_b)$ for a
system with $N_f=3$ and $N_b=1$ considering SM with $k=0.32 pN/nm$
(panel a) and MFM (panel b). Panel c shows the probabilities of the
states $(2,1)$ and $(3,1)$ as functions of $k$ within the SM. In all
the cases, the normalization of the probabilities is such that their
sum over the eight possible states is equal to one.}
\end{figure}

Now we study the probabilities of the different states $(n_f,n_b)$
for fixed $N_f$ and $N_b$. This is relevant for finding out to what
extent forward and backward motors coexist linked to the
microtubule, and for evaluating the mean number of engaged motors.
Let us take the case $N_f=3$, $N_b=1$ with $k=0.32 pN/nm$ as an
example. Figures \ref{pnfnb}.a and \ref{pnfnb}.b show the
probabilities of the different states $(n_f,n_b)$ for SM and MFM
respectively. Note that the state $n_f=n_b=0$ has null probability
as it determines the end of the simulation. The results clearly
indicate that the states with one engaged backward motor are much
more likely in the SM than in the MFM. Note that for the MFM, the
states with $n_b=0$ accumulate around a $85\%$ of the probability.
This means that the backward motor is detached most of the time. In
particular, the states $(3,0)$ and $(2,0)$ alone dominate the
dynamics a $80\%$ of the time. In contrast, the SM predicts that
the backward motor will be essentially half of the time engaged to
the microtubule. Moreover the probabilities for the states $(3,0)$,
$(3,1)$, $(2,0)$ and $(2,1)$ are all similar to each other. States with
the backward motor engaged are more likely in the SM than in the MFM
due to that, within the SM, the backward motor is unloaded a non
negligible part of the time and, when it is loaded, it is in a tug
of war only with those forward motors that are beyond the limit of
$110 nm$ from cargo. In contrast, in the MFM the engaged backward
motor is all the time in a tug of war with $N_f$ motors, each of
which is less loaded than the backward motor. Within the SM, the
probabilities of states with a backward motor engaged is found to
decrease with the stiffness $k$, as we show in Figure \ref{pnfnb}.c.
This is reasonable, since smaller stiffness lead to lower
probability of detachment and results in more permissive of tug of war
states. In the case of the system with $N_f=3$ and $N_b=2$ the
results (not shown) are completely analogous to those for $N_f=3$
and $N_b=1$ in figure \ref{pnfnb}. Thus, we can state generally that
the probabilities of states with engaged backward motors increase
when passing from MFM to SM and, within the SM, they increase with
decreasing $k$. This explains the increase of the mean number of
engaged motors when changing from MFM to SM and when decreasing $k$
(figure \ref{vyrN3}.c) as a consequence of the contribution of the
tug of war states, which have larger total number of motors than
single species states.

\begin{figure}[h!]
\centering
\resizebox{\columnwidth}{!}{\includegraphics{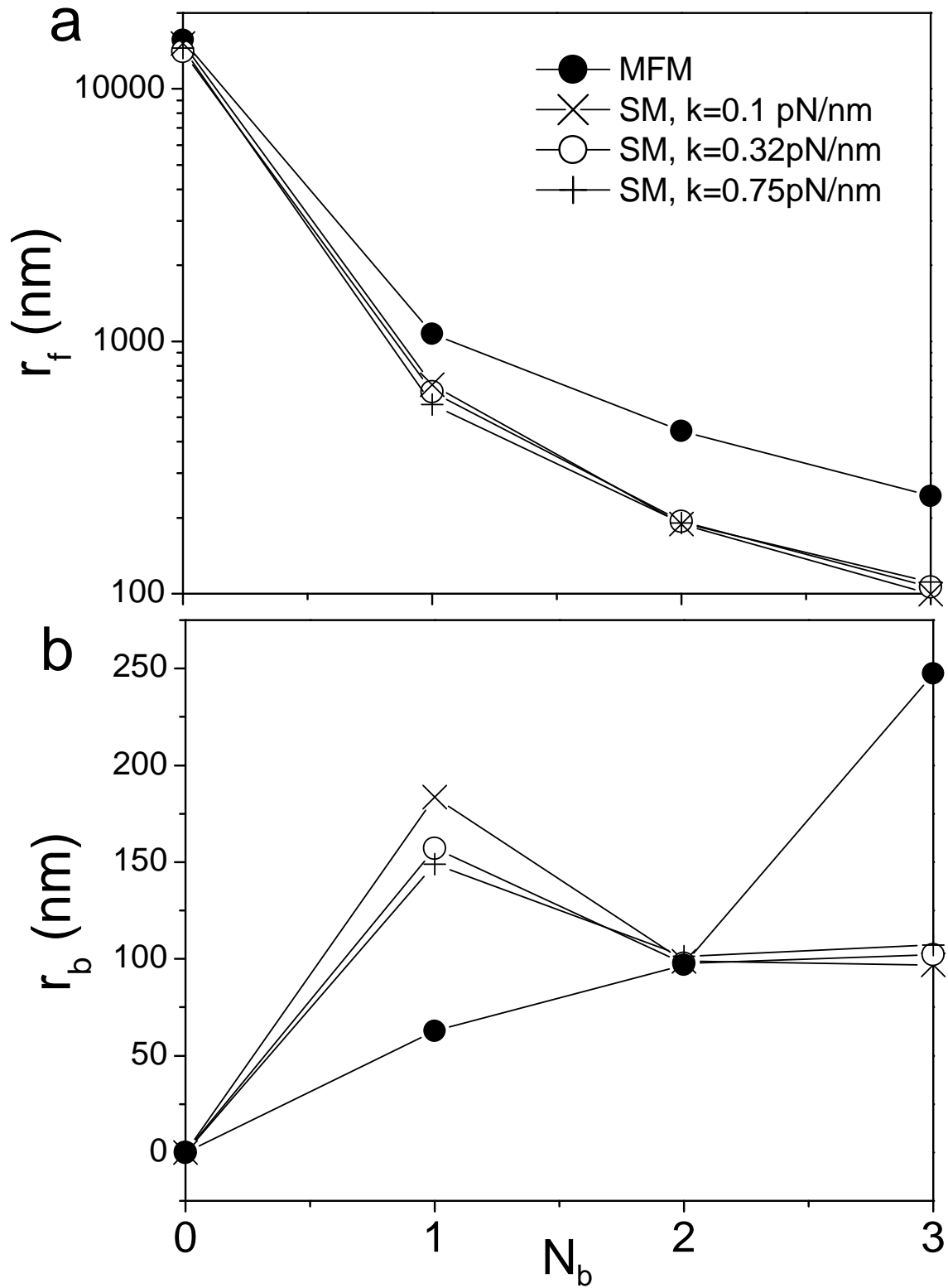}}
\caption{\label{rmasrmenos} Results for $r_f$ and $r_b$ as functions
of $N_b$ for the same systems analyzed in figure \ref{vyrN3},
considering MFM and SM.}
\end{figure}

In figure \ref{rmasrmenos} we show the results for $r_f$ and $r_b$
corresponding to the same systems analyzed in figure \ref{vyrN3}. As
expected, $r_f$ decreases monotonously with $N_b$ in both models,
since the probability of reversions grows with $N_b$. We also see
that the SM predicts shorter forward excursions than the MFM. This
is because the larger probability of having a backward motor engaged
leads to a larger probability of changing from forward to backward
motion. The results for $r_b$ in figure \ref{rmasrmenos}.b are much
more intriguing due to the abrupt variations with $N_b$.
Nevertheless, we can give an almost complete explanation for them.
Concerning the results for the SM, the counterintuitive fact that
$r_b$ is larger for $N_b=1$ than for $N_b=2$ is due to the
contribution of the state $(n_f,n_b)=(1,2)$ in the latter case. In
fact, this state is found to be the one that largely contributes to
$r_b$ for $N_b=2$ (it has probability $p=0.096$ while states $(0,2)$
and $(0,1)$ have only $p=0.035$ and $p=0.017$ respectively).
Clearly, the state $(1,2)$ is expected to produce smaller velocities
(and thus shorter runs) than state $(0,1)$ which is the only one
contributing to $r_b$ for $N_b=1$ (with $p=0.019$). However, note
that, for $N_b=2$, the accumulated probability of having $n_f>n_b$
is larger than for $N_b=1$. It thus happens that, for $N_b=2$,
backward excursions are shorter but more frequent than for $N_b=1$.
The results for $r_b$ in the MFM are less intriguing, except
maybe for the abrupt decrease when passing from $N_b=3$ to $N_b=2$.
This is simply related to that, due to equal force sharing, leaving
the symmetric situation $N_b=N_f$ works largely against the species
which results with lower number of motors.

\subsection{Influence of the viscous drag}

\begin{figure}[h!]
\centering
\resizebox{\columnwidth}{!}{\includegraphics{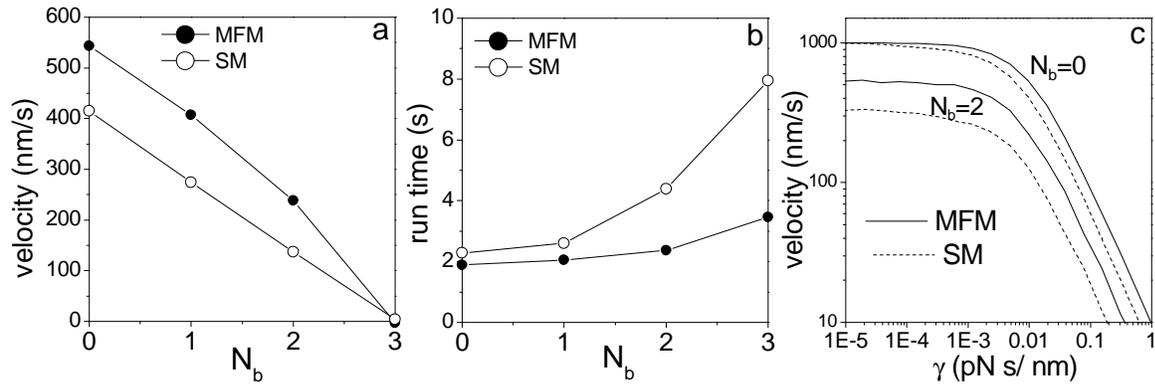}}
\caption{\label{fviscous} Influence of the viscous drag. Cargo mean
velocity (a) and run time (b) as functions of $N_b$ for fixed
$N_f=3$ considering $\gamma=9.42 10^{-3} pN s/nm$ for both models.
This value of $\gamma$ corresponds to a thousand times water
viscosity and a cargo of 500 nm radio. c) Cargo mean velocity as a
function of $\gamma$ for $N_f=3$ and different values of $N_b$ for
both models. All calculations within SM are for $k=0.32 pN/nm$.}
\end{figure}

Now we analyze bidirectional transport under non negligible viscous
drag. Figure \ref{fviscous}.a shows the cargo mean velocity as a
function of $N_b$ for $N_f=3$ considering a viscous drag equal to
$1000$ times that of water, both for MFM and SM. As expected, the
velocities are smaller than those for water viscosity shown in
figure \ref{vyrN3}.a (typically by a factor 1/2). The general
behavior of both models is similar to that for water viscosity, with
one relevant difference. Now, MFM and SM give different results even
for $N_b=0$. This is because, while in the MFM all the forward
motors share the load coming from the viscous drag, in the SM the
load acts essentially only on the pulling motors. The same would
occur when considering any other kind of external load force acting
against the advance of the cargo, as the results in \cite{mogilner}
for a single motor species suggest. Figure \ref{fviscous}.b shows
the run times for the same systems analyzed in Figure
\ref{fviscous}.a. As in the case of the velocities, we find that the
differences between both models extend to the case $N_b=0$. Finally,
we complete our analysis of the influence of viscous drag with the
results in figure \ref{fviscous}.c, which show the dependence of the
cargo velocity on $\gamma$ for systems with and without backward
motors. It can be seen that, although the general dependence on
$\gamma$ for all systems and models are similar to each other ($\tau_r$
is constant for $\gamma\lesssim 5\times 10^{-3} pN s/nm$ and
decreases exponentially for $\gamma\gtrsim 5\times 10^{-3} pN
s/nm$), the predictions of both models coincide only in the case of
low viscosity and no backward motors. Moreover, the differences
between the results from both models increase with the addition of
backward motors.

\section{Results for uneven forward and backward motors}

Now we leave the symmetric case and study the models considering
backward motor parameters that can be associated to cytoplasmic
dynein. For forward motors we continue using the kinesin-1
parameters considered in the previous section. Since the walking and
detachment properties of dyneins are not as well known as for
kinesin-1, the parameters for dyneins are not quite clear. Here we
consider two different sets of parameter's values (named simply as
set A and set B) based on the two proposals in
\cite{tugofwar,Muller,lipowskiBioPJ}. Set A has been considered
within MFM in \cite{tugofwar} and with small changes in
\cite{lipowskiBioPJ} on the base of previous experimental and
theoretical results (see Table 1 in \cite{tugofwar}, supporting
material in \cite{lipowskiBioPJ} and references therein). Set B was
obtained by fitting experimental data on Drosophila lipid-droplet
transport \cite{tugofwar} and is consistent with previous
experimental data \cite{setB}. According to set A, the main
differences between dyneins and kinesin-1 appear in the binding and
unbinding rates. In contrast, set B considers also important
differences on the typical velocities and stall forces of both motor
types. It is interesting thus to investigate the bidirectional
motion of cargo transported with kinesin-1 motors and both models of
dyneins.

Note that for the SM, in addition to binding, unbinding and velocity
parameters, it is also necessary to specify the forward-backward
ratio of jumps as a function of the load force. Since we have not
experimental data for dyneins, in this work we consider the same
exponential form used for kinesins. The use of any other reasonably
formula is not expected to produce relevant changes in the results.
In fact in \cite{zgz} it was shown that even the consideration of no
backward steps produces relatively small changes for the case of
transport by kinesins.

\begin{figure}[h!]
\centering
\resizebox{\columnwidth}{!}{\includegraphics{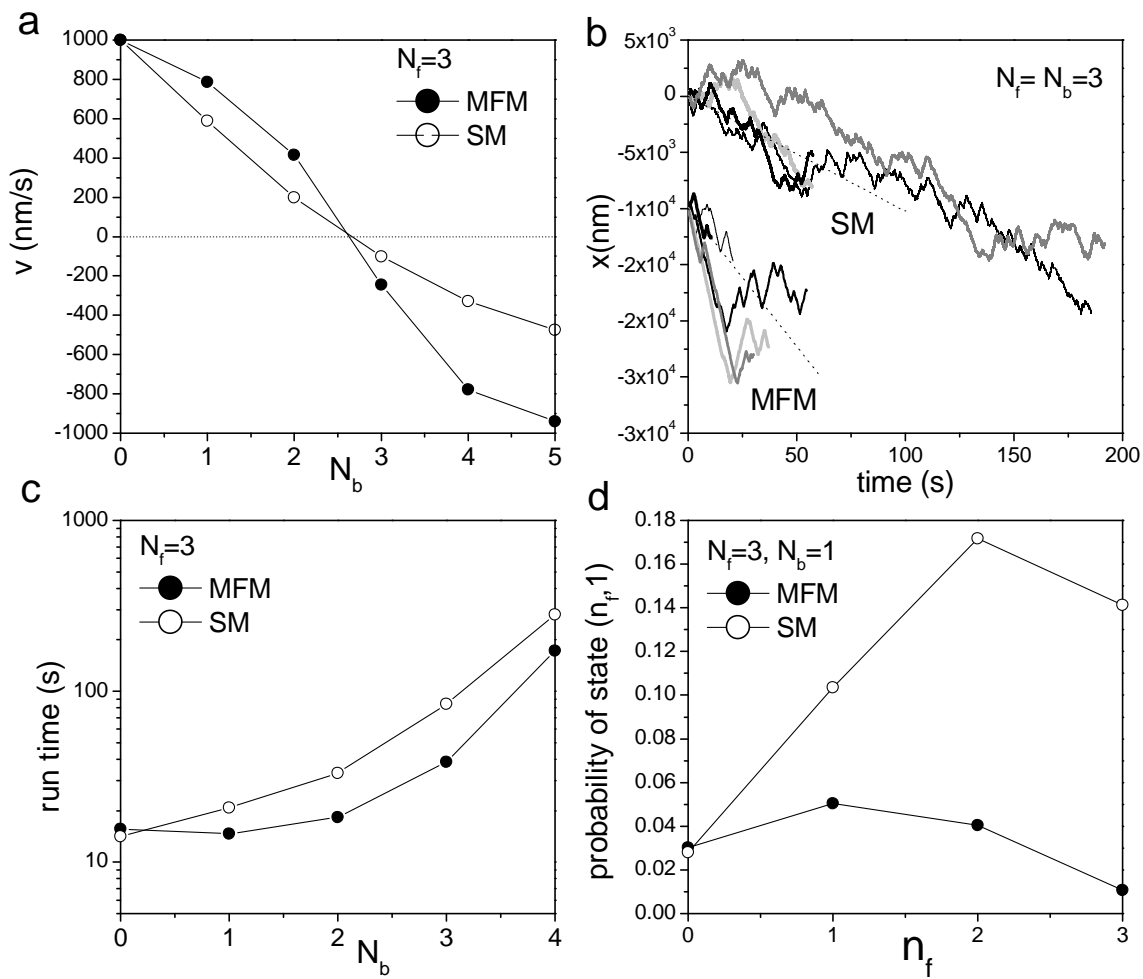}}
\caption{\label{fkindinA} Kinesins and dyneins. Cargo mean velocity
(a), cargo trajectories (b), mean run time (c) and probability of
states with $n_b=1$ (d) for a cargo pulled by kinesin-1 and dynein
considering set A parameters for dyneins and negligible viscosity
conditions for both models. Set A: $F_{sb}=7. pN, F{db}=3.18 pN,
\epsilon_b=0.27 s^{-1}, \Pi_b=1.6 s^{-1}, v_{0b}=1000 nm/s$ and
$v_{1b}=6 nm/s$.}
\end{figure}

In figure \ref{fkindinA} we show results for cargo transport by
kinesins-1 and set A dyneins under negligible viscous drag. Since
the differences between the parameters for both types of motors are
relatively small, the results are similar to those for symmetrical
motors. The effects of the asymmetry are mainly notable in the case
$N_f=N_b=3$, for which we observe a net backward motion. This can be
seen both in the velocities in figure \ref{fkindinA}.a and in the
trajectories in figure \ref{fkindinA}.b. The fact that set A dyneins
win the tug of war for $N_f=N_b$ is mainly due to their slightly
larger stall force. The differences between the predictions of MFM
and SM for the velocities (figure \ref{fkindinA}.a) and the run
times (figure \ref{fkindinA}.c) are considerably relevant, and they
occur in a similar fashion to that observed for symmetrical motors.
The same happens with the probabilities of having a backward motor
engaged (figure \ref{fkindinA}.d), which are found to be larger for
the SM than for the MFM. As in the case of symmetrical motors, this
latter result helps us to understand the differences in the run
times from both models. Going back to figure \ref{fkindinA}.b, we
see that the SM trajectories are much more winding than those from
MFM. This phenomenon is related to the reduction of $r_f$ analyzed in
the case of symmetric motors. The larger rate of reversion in the SM
is due to that only some of the engaged motors pull the cargo at a
given time and, in addition, it is more likely to have opposing
motors engaged than in the MFM.

\begin{figure}[h!]
\centering
\resizebox{\columnwidth}{!}{\includegraphics{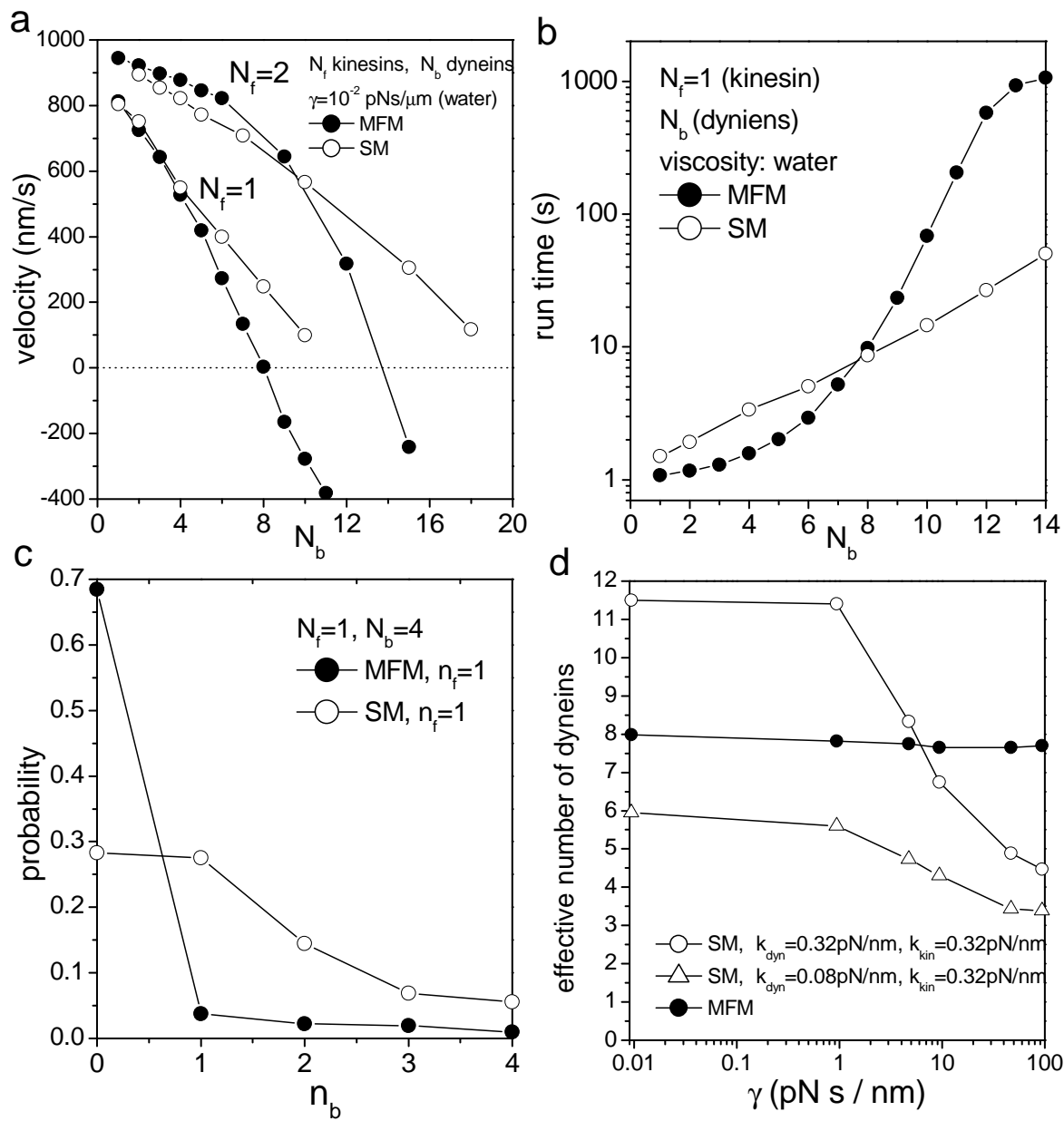}}
\caption{\label{fkindinB} Kinesins and dyneins. Cargo mean velocity
(a), mean run time (b) probability of different states (c) and
effective number of dyneins leading to vanishing mean cargo velocity
(d) for a cargo pulled by kinesin-1 and dynein considering set B
parameters for dyneins and negligible viscosity conditions for both
models. Set B: $F_{sb}=1.1 pN, F{db}=0.75 pN, \epsilon_b=0.27
s^{-1}, \Pi_b=1.6 s^{-1}, v_{0b}=650 nm/s$ and $v_{1b}=72 nm/s$.}
\end{figure}

Now we consider dynein with parameter set B. In this case, dynein is
a motor sensibly weaker than kinesin-1, since it has much lower
stall force, much lower detachment force, lower attaching
probability and also lower ratio $F_s/F_d$. Thus, for equal number
of kinesin and dyneins, the kinesins win the tug of war. In fact, we
find that several dyneins are needed to produce average null
velocity for a cargo pulled by only one kinesin. In figure
\ref{fkindinB}.a we show the cargo mean velocity as a function of
$N_b$ for systems with $N_f=1$ and $N_f=2$. It can be seen that for
small $N_b$ the velocity is similar for both models (almost
coincident for the case $N_f=1$). In contrast, for relatively large
values of $N_b$ the predictions of both models differ substantially.
In particular, the number of dyneins needed to attain zero average
velocity for a cargo pulled by one or two kinesins are considerably
different. For $N_f=1$ we find $N_b\sim 8$ for the MFM and $N_b\sim
12$ for SM, while for $N_f=2$ we find respectively $N_b\sim 14$ and
$N_b\sim 20$.

Figure \ref{fkindinB}.b shows the dependence of the run time on the
number of dyneins for a cargo pulled by one kinesin. It can be seen
that, even for $N_b<4$ for which both models give similar mean
velocities, they predict quite different results for $\tau_r$. In
the region $n_b \sim 8$, the situation is the opposite, both models
give similar run times but they predict quite different velocities.
For $n_b > 8$ the run time in the MFM grows very fast with $n_b$ due
to that dyneins win the tug of war (the mean velocity is negative).
This makes dyneins to remain mostly attached while the only kinesin
detaches. Thus, the total number of engaged motors which controls
the run time increases.

In figure \ref{fkindinB}.c we show results for the probabilities of
the different states for a system with $N_f=1$ and $N_b=4$. For
simplicity, we show only the probabilities of states with $n_f=1$
(states with $n_f=0$ have very small contributions in both models).
Again, the SM gives a much larger probability of engagement of
backward motors than the MFM.

Finally, we briefly explore the influence of the viscosity and of
possible differences on the stiffness of motors from both species.
In figure \ref{fkindinB}.d we study the effective number of set B
dyneins needed to achieve zero mean velocity when pulling against
one kinesin. We consider different values of $\gamma$ and $k_b$.
Note that our results provide non integer effective values for $N_b$
which correspond to interpolations leading to zero cargo velocity.
We see that for the MFM the results are almost independent of
$\gamma$ at a value close to $N_b=8$. Interestingly, this value can
be estimated by equating the powers produced by both motor teams
considering all the motors attached at stall force. Namely,
considering $N_b\times v_{sb}\times f_{sb}=v_{sf}\times f_{sf}$ we
get to $N_b=8.39$. The behavior within the SM is much more complex.
The effective number of dyneins needed to stop a kinesin depends
both on the viscosity and the stiffness. Actually, it decreases with
$\gamma$ and increases with $k_b$. The decrease with $\gamma$ is
clearly due to that viscosity helps to stop the cargo. The
increase with $k_b$ is due to that larger $k_b$ lead to less force
production by dyneins, due to easier detachment. Note that we have
have considered a quite small value for the dynein's stiffness
($k_b=0.08 pN/nm$). The reason for this is twofold. First, it is the
only way to reduce to reasonable values the effective number of set
B dyneins needed to attain null velocity when pulling against one
kinesin. Second, recent experiments \cite{Bruno} for transport
mediated by dynein and kinesin-2 reported values of the stiffness in
such range.

\section{Conclusions}

Cargo transport along microtubules mediated by two opposing motor species provides interesting
challenges both from the experimental and theoretical points of view. With the main aim of
understanding the consequences of specific modeling assumptions, in this paper we have theoretically
analyzed several aspects of the problem considering two different mathematical models: the mean field
model (MFM) \cite{tugofwar}, and a recently introduced stochastic model (SM) \cite{zgz} which share
some commons with models in \cite{grossCB2008} and \cite{mogilner}. The main difference between the
MFM and the SM stems from the assumptions of force sharing by the different motors. The MFM assumes
equal load sharing by all the engaged motors of the same polarity while SM considers individual cargo
motor linking allowing for uneven force sharing.

Our main results indicate that both models show complete agreement only when there is essentially no
load to share, that is, in systems with a single type of motors and with no relevant viscous effects.
In other situations, the MFM predicts larger cargo mean velocity and smaller mean run time than the
SM. We have found that the differences in the velocities are mainly due to the fact that, within the
SM (and in agreement with statements in \cite{grossCB2008,mogilner,Jamison}), only some of the
engaged motors pull the cargo at a given time. Moreover, the probability of engaged backward motors
during forward excursions is larger in the SM tan in the MFM. This leads also to a larger rate of
reversions within SM when compared with the MFM, or equivalently, to shorter excursions toward each
polarity. The difference between the mean velocities predicted by both models is found to increase
with the viscosity. We have also found that the mean run time is essentially controlled by the mean
number of engaged motors at a given time, which depends on the probabilities of the different
motility states and, ultimately, on the force sharing assumptions. Our results for the SM show that
the mean cargo velocity turns out to be rather independent of the stiffness of the motor-cargo link ($k$). In
contrast, the mean run time decreases with $k$ approaching the MFM results for large $k$.

These conclusions were obtained analyzing ideal systems in which motors of opposite polarities have
identical dynamical properties. In addition, in section 4 we have provided results for the asymmetric
case of transport driven by kinesin-1 and cytoplasmic dyneins, considering two different sets of
parameters for dyneins usually found in the literature.

Finally, it is interesting to mention the possibility of considering a hybrid-modeling framework
taking advantage of the benefits of both kinds of models: the simplicity and computational advantage
of the Gillespie formulation of the MFM, and the higher reliability of models including uneven force
sharing. Note that the SM provides an alternative way to compute (numerically) the transition rates
and velocities of the motility states entering in the Gillespie algorithm considering uneven load
sharing. The same could be done with models such as those in references \cite{grossCB2008,mogilner}.
Thus, when fitting the single motor parameters needed to reproduce experimental trajectories,
different intermediate procedures combining computations with both kinds of models could be imagined,
depending on the particular problem and on the a priori knowledge of the parameters.

\section*{Acknowledgments}
SB acknowledge L. Bruno for useful discussions. Financial support from the Argentinean agencies
CONICET (under Grant No. PIP 11220080100076) and CNEA, and from of the Spanish MICINN through project
FIS2008-01240, co-financed by FEDER, are also acknowledged.
\section{Figures}

\end{document}